\documentstyle[epsf,11pt]{article}
\begin {document}
\title{Towards Solving QCD -  \break
        The Transverse Zero Modes in Light-Cone Quantization.}
%---+----1----+----2----+----3----+----4----+----5----+----6----+----7----+
\author{Hans-Christian Pauli and Alex C. Kalloniatis\\
Max-Planck-Institut f\"ur Kernphysik \\
D-69029 Heidelberg  \\
and  \\
Stephen~S. Pinsky \\
Physics Department, The Ohio State University \\
Columbus, OH 43210 \\ }
\date{17 March 1995}
\maketitle

\begin{abstract}
We formulate QCD in (d+1) dimensions using
Dirac's front form with periodic boundary
conditions, that is, within Discretized Light-Cone Quantization.
The formalism is worked out in detail
for SU(2) pure glue theory in (2+1) dimensions
which is approximated by restriction to the lowest
{\it transverse} momentum gluons.
The dimensionally-reduced theory turns out to be
SU(2) gauge theory coupled to adjoint scalar matter in
(1+1) dimensions. The scalar field is the
remnant of the transverse gluon.
This field has modes of both non-zero
and zero {\it longitudinal} momentum. We categorize
the types of zero modes that occur into
three classes,
dynamical, topological, and constrained,
each well known
in separate contexts.
The equation for the constrained mode is explicitly worked out.
The Gauss law is rather simply resolved to extract
physical, namely color singlet states.
The topological gauge mode is treated
according to two alternative scenarios related to the
elimination of the cutoff.
In the one, a spectrum is found consistent with pure
SU(2) gluons in (1+1) dimensions.
In the other, the gauge mode excitations
are estimated and their role in the spectrum with
genuine Fock excitations is explored.
A color singlet state is given which satisfies Gauss' law.
Its invariant mass is estimated and discussed in the physical limit.
\end{abstract}
\newpage
\section{Introduction}
%---+----1----+----2----+----3----+----4----+----5----+----6----+----7----+
It remains notoriously difficult
to understand the low-energy
regime of Quantum Chromodynamics (QCD) in terms of the simplistic
but otherwise successful constituent quark picture.
In line with the formulation of Feynman's parton model \cite{Fey69}
in the infinite momentum frame
\cite{Wei66,LeB80}, a promising approach could be that of Pauli and Brodsky
\cite{PaB85a,PaB85b}
which adapts Dirac's `front form' Hamiltonian dynamics \cite{Dir49}
for nonperturbative
quantum field theory. Perhaps a misnomer, the method carries the name
Discretized Light-Cone Quantization (DLCQ).
Numerous applications have been carried out with reasonable success
in extracting bound state spectra and wavefunctions
for both (1+1) dimensions in Abelian \cite{EPB87,ElP89,Hil91}
and non-Abelian gauge theories \cite{Hor90,DKB93,BDK93,HeK94}
as well as for higher space-time dimensions \cite{Hol92,KPW92,WiH93,THY94}.
%The idea was later
%taken up by Wilson and collaborators in the related
%Light-Front Tamm-Dancoff approach \cite{PHW90}.
DLCQ combines the aspects of a simple vacuum
\cite{Wei66} with a careful treatment of the infrared
degrees of freedom. The latter are controlled by the finite
volume regularization
of the method. There is legitimate hope that one can thus get both
a manageable treatment of the `vacuum problem'
and explicit invariant mass spectra and
wavefunctions for physical particles.

%---+----1----+----2----+----3----+----4----+----5----+----6----+----7
In this paper we concern ourselves with more than just the vacuum
problem, which in this context manifests itself as the so-called
`zero mode problem'.
Contrary to the original expectations of the framers of DLCQ, the
zero momentum modes of the Lagrangian field operators have proved
far more than just a ``set of measure zero''.
Indeed, in the $\phi^4$ theory in
(1+1) dimensions they are crucial in reproducing the
vacuum properties of the theory, namely spontaneous
symmetry breaking and a vacuum condensate
\cite{Rob93,Hei91,Hei92a,Hei92b,BPV93,PiV94a,HPV94}.
This is achieved in the DLCQ framework by the property of the zero
mode of $\phi$ being, not dynamical, but satisfying a
constraint \cite{MaY76}.
This preserves the simplicity of the vacuum and
thus the partonic picture of light-cone field theory.
The desired symmetry is {\it explicitly} broken upon solution of the
constraint by non-perturbative approximation and substitution in the
Hamiltonian \cite{BPV93,PiV94a,HPV94}.
Now the four-point coupling of {\it gluons} suggests at least some
aspect of this feature will
be present in QCD(3+1). We thus set out in the sequence of papers
\cite{KaP93,KaP94,KaR94,KPP94} to {\it disentangle} the constraint
problem from that of the gauge-symmetry of the non-Abelian theory.
Constrained zero modes occur even in Abelian theory
\cite{KaP93,KaP94} for which one requires gauge-fixing to
solve \cite{KaR94} and that gauge choice may {\it not} in general be that
on the non-zero modes.
But not all zero modes in a gauge theory are constrained.
Less significant in QED(3+1),
some of these dynamical modes are intimately related to
the non-trivial topology of both the hyper-torus implicit in DLCQ and
become important in the presence of
non-Abelian gauge groups \cite{KPP94}.
We thus find a diverse range of zero mode types, all of which will
evidently be present in QCD(3+1).
Treating these types {\it together} in a single theory,
albeit a still simplified one, is
the subject of this paper.

The essential principle we shall use for getting
to simpler theories from QCD(3+1)
was espoused in \cite{KPP94}.
Lower dimensional `regimes' of a higher dimensional theory
can be systematically explored because of the finite volume regularization:
zero and non-zero momentum modes can be cleanly distinguished and so
one can, for example, excite one and not the other.
One thus obtains effective theories in lower space-time dimensions
that are not identical to the original theory defined {\it a priori}
in lower dimensions. This is essentially `dimensional reduction'
\cite{Sie79,Sie80}.
A similar idea for the instant form was recently suggested by
\cite{FeJ95}.
In \cite{KPP94} we examined (1+1) dimensional pure SU(2) gauge theory
coupled to external sources in DLCQ.
We suppressed {\it all} momentum excitations and obtained
a (0+1) dimensional -- quantum mechanical -- theory of a single
gluon zero mode whose dynamics depended on external sources coupled
to the gluons. This mode corresponds to the quantized
flux loop around the circle defining space. The mode is purely
of topological origin, and thus the theory was manifestly isomorphic
to a quantization in the instant form defined on the analogous
topology \cite{Het93a,Het93b}. This is one of the types of zero modes
discussed above.

The `next step', taken in this paper, is to begin with
(2+1) dimensional SU(2) theory and looking at the nested (1+1) dimensional
theory by suppressing transverse gluon momentum excitations.
The topological
mode appears here again, but now coupled to true dynamical field modes
that are the Fock modes of the transverse gluon component. As well,
other types of zero modes appear. This is now the simplest type
of non-Abelian gauge theory we can construct in which all the types of
zero modes encountered previously couple together into a
nontrivial dynamical problem. Many of the structures we unveil here
were already foreseen in 1981 by Franke et al. \cite{FNP81,FNP82}
in (3+1) dimensions.
It will become clear in the course of the present work that
these types of models, first discussed in DLCQ by \cite{DKB93,BDK93}
but without zero modes and assuming only color singlet string states,
enable insight into how to overcome the obstacles that impeded Franke.

The aim of this work then is two-fold.
The first is a formal aim: to show how a treatment of the non-Abelian
gauge theory can be achieved which keeps the advantages of the
front-form approach while controlling the infrared problem.
This we succeed in doing insofar as we can give a Hamiltonian
in which the nature of all modes --- zero and non-zero --- is
clarified and their means of solution at least understood.
Sections 2 to 4 deal with these formal aspects.
In particular, in Section 3 we make the restriction in
the gauge fixing to the so-called `fundamental modular domain'
\cite{vBa92}.
The second aim is to gain insight into the physical spectrum of the
aforementioned Hamiltonian of the pure glue theory by at
least semi-analytic methods. Here we relax rigour and
make several simplifying assumptions within the context of a cutoff
regularization of the large momentum region of the theory.
A point-splitting treatment will be presented
elsewhere \cite{PKP95}. Several insights
into the spectrum are obtained. The purely contraction
parts of the Hamiltonian lead to the potential for the Schr\"odinger
equation in the gauge zero mode sector. This is the analogue for
the problem we solved in \cite{KPP94}. Two alternative methods are
described for dealing with the singular structure of the potential,
either keeping the cutoff or `renormalizing' the potential.
In both scenarios, we are able to simultaneously diagonalize the
energy and momentum operators.
This is discussed in Section 5.
However, in the absence of a definite
counterterm for the renormalization approach we use the
solution to the gauge zero mode keeping the cutoff and implement it
in the particle sector of the theory.
We give a color singlet state which is
an eigenstate to part of the Hamiltonian.
The invariant mass squared of this state is seen to diverge when
all cutoffs are taken to their physical values. We comment on this in
the discussion.

%---+----1----+----2----+----3----+----4----+----5----+----6----+----7----+
\section{Formulation for Pure SU(N) Gauge Theory.}
Consider an SU(N) gauge theory
without fermions in $d+1$ dimensions defined by
the Lagrangian density
\begin {equation}
    {\cal L}  = - {1\over2}
                {\rm Tr} \bigl({\bf F}^{\mu\nu} {\bf F}_{\mu \nu} \bigr)
 \ , \ {\rm with}\quad
    {\bf F}^{\mu\nu} \equiv \partial^\mu  {\bf A}^\nu
   - \partial^\nu  {\bf A}^\mu
   + i g \bigl[ {\bf A}^\mu, {\bf A}^\nu\bigr] \
 \equiv \partial^\mu  {\bf A}^\nu
  - {\bf D}^\nu  {\bf A}^\mu
\ .
\label{Lagr}
\end {equation}
The $ {\bf A}^\mu$ are the SU(N) {\it vector potentials}.
We shall reserve the term `gauge potential' for something
else, discussed below.
The energy-momentum tensor is derived from Eq.(\ref{Lagr})
in the usual way \cite{JaM80} by
$  \Theta^{\mu\nu} = 2{\rm Tr} \bigl({\bf F}^{\mu\kappa}
   {\bf F}_\kappa^{\phantom{\kappa}\nu}\bigr)
 - g^{\mu\nu} {\cal L}  $.
This and the notation are explained in more detail in
Appendix \ref{sec:notapp}. But at this stage we keep the discussion
general for SU(N) for arbitrary number of
colors $N$.
In the {\it front form}, it is convenient to separate
the Lorentz indices $\mu(\nu)$
into {\it longitudinal} values $\alpha(\beta)= +,-$ and
{\it transversal} values $j(i)= 2,3,\dots ,d$.
The indicial sums in the Lagrangian and the (light-cone)
energy-density $ \Theta^{+-}$
then disentangle cleanly,
\begin {equation}
    {\cal L}   = - {1\over2} {\rm Tr} \bigl(
    {\bf F}^{\alpha\beta} {\bf F}_{\alpha\beta}
  + {\bf F}^{ij} {\bf F}_{ij}
  + 2  {\bf F}^{\alpha j} {\bf F}_{\alpha j}  \bigr)
 \quad \  {\rm and} \quad
 \Theta^{+-} = {1\over2}{\rm Tr} \bigl(
   {\bf F}^{\alpha\beta} {\bf F}_{\beta\alpha}
 + {\bf F}^{ij} {\bf F}_{ij}       \bigr) \ ,
\end   {equation}
respectively.
%---+----1----+----2----+----3----+----4----+----5----+----6----+----7
The dimensionality of the problem is not manifest but
resides in the dimensions of the fields. Working in dimensions of
length $l$,
we have for the field and coupling the usual
$dim[ {\bf A}^\mu] \equiv l^{(1-d)/2}$
and $dim[ g ] \equiv l^{(d-3)/2}$.
The energy-density in (3+1)
and (2+1) dimensions has the simple structure
\begin {equation}
   \Theta^{+-}_{3+1} = {\rm Tr} \bigl(
     {\bf F}^{-+} {\bf F}^{-+} +
     {\bf F}^{23} {\bf F}^{23} \bigr)         \quad\  {\rm and}\ %
   \Theta^{+-}_{2+1} = {\rm Tr} \bigl(
     {\bf F}^{-+} {\bf F}^{-+} \bigr) \ .
\end   {equation}
As in previous work \cite{KaP93,KaP94} it is convenient to
disentangle such an expression into `zero modes' and `normal modes'.
A {\it zero mode} of some function
$f(x^-, x_\bot)$
with respect to any one of the space coordinates, say
$ y $ with interval length $L_y$, is
defined by
\begin {equation}
  < f(\bar{x})  >_o \equiv {1\over{2L_y}}
  \int_{-L_y}^{L_y} dy \, f(\bar{x},y) \ ,
\end {equation}
where $\bar{x}$ are the remaining spatial coordinates not being
integrated over.
When $ y $ is the longitudinal direction
$ x^-$ we shall denote the zero mode by
${\hbox{\vbox{\ialign{#\crcr
    ${~\scriptstyle \circ\,}$\crcr
   \noalign{\kern1pt\nointerlineskip}
    $\displaystyle{f}\hfil$\crcr}}}}$.
The {\it normal mode} is, in general, the complement
${\hbox{\vbox{\ialign{#\crcr
    ${~\scriptstyle n\,}$\crcr
   \noalign{\kern1pt\nointerlineskip}
    $\displaystyle{f}\hfil$\crcr}}}}
 = f - < f  >_o $.
We see then that since $ P^-$ involves an integration
of the density with respect to the spatial coordinates $x^i$,
it is like
evaluating the transversal `zero mode' of the energy density.
Taking the zero mode with respect to any
space-like coordinate, one realizes that the Hamiltonian
is {\it additive}
in the zero and normal mode contributions, {\it i.e.}\
\begin {equation}
   {\rm Tr}  <  {\bf F}^{-+} {\bf F}^{-+}  >_o \equiv
   {\rm Tr} \bigl(< {{\bf F}}^{-+}  >_o
                < {{\bf F}}^{-+}  >_o \bigr) +
   {\rm Tr}  <
 {\hbox{\vbox{\ialign{#\crcr
    ${\,\scriptstyle n\,\,}$\crcr
   \noalign{\kern1pt\nointerlineskip}
    $\displaystyle{{\bf F}}^{-+}$\crcr}}}}
 {\hbox{\vbox{\ialign{#\crcr
    ${\,\scriptstyle n\,\,}$\crcr
   \noalign{\kern1pt\nointerlineskip}
    $\displaystyle{{\bf F}}^{-+}$\crcr}}}} >_o \ .
\end   {equation}
Of course one should not take this too far, since zero and
normal modes of the individual quantum fields can (and do)
reside in both terms of this expression. However the
separation allows for a conceptual simplification:
By lack of insight, the original formulation of
Discretized Light-Cone Quantization (DLCQ)
was formulated in terms
of {\it only} the normal modes.
It should be useful and even complementary
to analyze the theory in terms of only the transversal zero modes.

%{\it The formulation of the model.}
We therefore consider a model which only has
{\it transversal zero modes}
by requiring
\begin {equation}
  \partial_i  {\bf A}^\mu = 0 \ , \; {\rm{for \; all}} \; \mu \ .
\label{restrict}
\end   {equation}
This is to be regarded as a genuine dynamical restriction on the full
theory. Of course this theory will now involve both zero and
normal mode longitudinal gluon excitations.
Since the lengths $ L_{\!\bot} $ and $ L $ are now decoupled in scale,
it is convenient to readjust units by scaling out the transverse
length $ L_{\!\bot} $,
\begin {equation}
    {\bf A}^\mu \longrightarrow \widetilde  {\bf A}^\mu =
    {\bf A}^\mu (2 L_{\!\bot})^{(d-1)/2} \  {\rm and}\quad
    g      \longrightarrow \widetilde  g  =
    g  (2 L_{\!\bot})^{(1-d)/2} .
\label{transcale}
\end   {equation}
In the sequel the tilde is dropped. The dimensionality resides
then only in the Lorentz indices.
For simplicity we now restrict ourselves to consider the original
theory has having beeen formulated in (2+1) dimensions.
The result of the assumption Eq.(\ref{restrict}) is
to have {\it dimensionally reduced} the (2+1) theory to (1+1) dimensions.
However a reminder of the original (2+1) structure resides in the
continued presence of the transverse gluon component $ {\bf A}^1$.
In the spirit of Siegel \cite{Sie79,Sie80}, who introduced dimensional
reduction for regularizing supersymmetric theories,
we identify this gluon as a {\it scalar} field
$ \Phi $ transforming under the
adjoint representation of the color group. To completely avoid
playing with Lorentz indices we go further and introduce the notation
$$
%\begin {equation}
     {\bf A}^\mu = ({\bf A}^+, {\bf A}^-, {\bf A}^1) \equiv
    ({\bf V} , {\bf  A} , \Phi).
$$
The model-theory then takes the form of a (1+1) dimensional
non-Abelian gauge theory covariantly coupled to scalar adjoint matter
\begin {equation}
    {\cal L}  = {\rm Tr} \Bigl(
        - {1\over2} {\bf F}^{\alpha\beta} {\bf F}_{\alpha \beta}
       + {\bf D}^\alpha \Phi   {\bf D}_\alpha \Phi \Bigr) \ .
\end   {equation}
The covariant derivative $ {\bf D}_\alpha$ is implicitly defined in
Eq.(\ref{Lagr}).
A similar treatment of this theory in DLCQ was recently given by
\cite{DKB93,BDK93}.
The equations of motion in the two parts of the theory
can be deduced from the sourceless color Maxwell equations and are
%---+----1----+----2----+----3----+----4----+----5----+----6----+----7
\begin {equation}
   {\bf D}_\beta  {\bf F}^{\beta \alpha} =  g  {\bf J}_{\rm M}^\alpha
   \ , \ {\rm with}\quad
   {\bf J}_{\rm M}^\alpha = - i \bigl[ \Phi , {\bf D}^\alpha \Phi \bigr]
 \ ,\quad\  {\rm and} \quad
   {\bf D}^\alpha  {\bf D}_\alpha \Phi  = 0 \ .
\label{eqofmot}
\end    {equation}
Note that the `matter current' $ {\bf J}_{\rm M}^\alpha$ is not conserved,
$\partial_\alpha {\bf J}_{\rm M}^\alpha \neq 0$,
whereas the
total `gluon current'
$ {\bf J}^\alpha_{\rm G} =   {\bf J}_{\rm M}^\alpha
  -i\bigl[ {\bf F}^{\alpha\beta} , {\bf A}_\beta \bigr] $
is conserved.
To quantize the theory in terms of as few redundant
degrees of freedom as possible it is essential to fix the gauge.
We follow the procedure given in \cite{KPP94} and find that
$A^+$ only has a zero mode, i.e. $ \partial_-  {\bf A}^+ = 0$.
At this point we
specialize to SU(2). Then a single rotation in color space
suffices to diagonalize the SU(2) color matrix $ {\bf A}^+$.
The simple way to see this mode
cannot be gauged away is that it is related to the
Wilson loop for a contour (line) along the $x^-$
space: a gauge invariant quantity that cannot be set
to a fixed value by gauge choice. There remain a set of
`large' gauge transformations which generate shifts in
$ {\bf V} $ known as Gribov copies \cite{Gri78,Sin78}.
These matters are discussed further later and in
Appendix \ref{sec:GribFock}.
In the instant form this gauge has been used
by a range of authors,
\cite{Het93a,Het93b,LaS92,LSK93} to list a few.
In a context related to the front form it has also been
used by \cite{LTL91}.
Finally, the diagonal zero mode of $ {\bf A}^-(x^+_0)$ can be gauged
away \cite{KPP94} at some fixed light-cone time $x^+_0$.
For writing the Hamiltonian later, it is convenient to choose
this time as $x^+_0 = 0$, the null-plane initial value surface on which we
specify the independent fields.

In this gauge then,
$    {\bf F}^{-+} = \partial_+ {\bf V} - {\bf D}_- {\bf  A} $.
The first of our three equations of motion, $\alpha = +$, is
simply Gauss' law,
$  {\bf D}_-  {\bf F}^{-+} = - {\bf D}_-^2  {\bf  A} =  g
   {\bf J}_{\rm M}^+ $,
realized here as a {\it second class} constraint in the
nomenclature of Dirac \cite{Dir64,HRT76,Sun82}. In the
absence of gauge-fixing these are first class constraints.
They are a consequence of, and generate, gauge-symmetry.
With the gauge-fixing, these can be realized as
quantum operator constraints with an exception which we discuss
below. This aside, we can be cavalier and solve this equation
strongly yielding
\begin {equation}
    {\bf  A} = - g  {1\over   {\bf D}_-^2} {\bf J}_{\rm M}^+ \ .
\end    {equation}
The exception is a remnant first class constraint,
namely the zero mode diagonal part of Gauss' law.
The reason for this is that in selecting out the
color-three direction we have factored the
group SO(2)$\otimes$SO(2) from SU(2) (locally
isomorphic to SO(3)), leaving a subgroup of SO(2)
rotations which leave the preferred axis invariant.
This itself is isomorphic to U(1). In other words
we have a residual symmetry with respect to
global Abelian gauge transformations. It means in the
quantum theory there are redundant states in the Hilbert
space corresponding to the ``fibres'' or ``orbits'' in gauge
configuration space. Unique representatives on these orbits
must be selected by projecting a `reduced Hilbert space'
out of the larger space by requiring the selected, so-called
`physical' \cite{Dir64}, state vectors have vanishing `charge'
$  Q_0 \vert {\rm phys} \rangle = 0 $. The charge
here is the zero mode of the matter density
$ {\bf J}^{+}_{\rm M, diag} $.
Though this does not necessarily imply that $|{\rm{phys}}\rangle$
is an actual state in the physical {\it spectrum},
states outside this subspace render the theory
{\it a priori} inconsistent.
Analogous constraints can be found in many other
contexts, for example in \cite{MiP94}.

The second Yang-Mills equation of motion, with
$\alpha = -$, is
a genuine dynamical equation for $ {\bf V}  $, {\it i.e.}\
\begin {equation}
 \partial_+ \partial_+  {\bf V}
  + 2 g  i\bigl[ {\bf  A} , \partial_+  {\bf V} \bigr]
  -\partial_+ \partial_-  {\bf  A}
  -i g \bigl[ {\bf  A} , \partial_-  {\bf  A} \bigr]
  -i g \bigl[ {\bf V} , \partial_+  {\bf  A} \bigr]
  + g^2
 \bigl[ {\bf  A} ,\bigl[ {\bf V} , {\bf  A} \bigr] \bigr]
  = -  g   {\bf J}_{\rm M}^- \ .
\label{zeqm}
\end    {equation}
Since we are ultimately interested in a Hamiltonian
treatment of the dynamics of $ {\bf V} $,
we do not address ourselves to solving
dynamical Euler-Lagrange equations.

Next we discuss the equations in the scalar field sector,
$  {\bf D}^\alpha  {\bf D}_\alpha \Phi  = 0 $,
which reads explicitly
\begin {equation}
 \partial_- \bigl(\partial_+ +  {\bf D}_+ \bigr) \Phi
  + i g \bigl(\ \partial_+ \bigl[ {\bf V} , \Phi \bigr]
  + \bigl[ {\bf  A}^\alpha , {\bf D}_\alpha \Phi \bigr] \ \bigr)
 = 0\ . \label{phieqm}
\end    {equation}
The zero mode of its color-diagonal part
will turn out to be a true operator constraint equation,
occuring as second class in the Dirac procedure.

We complete this overview of the theory
with the Hamiltonian and light-cone momentum operator.
One calculates them to be, respectively
\begin {eqnarray}
    P^- & = & \int_{- L}^{+ L} \!\! dx^- {\rm Tr}
 \ \bigl(\partial_+ {\bf V} - {\bf D}_- {\bf  A} \bigr)^2
           = \int_{- L}^{+ L} \!\! dx^- {\rm Tr}
 \ \bigl(\ \partial_+ {\bf V} \partial_+ {\bf V}
  - g^2  {\bf J}_{\rm M}^+{1\over {\bf D}_-^2} {\bf J}_{\rm M}^+
 \ \bigr) \ ,
\label{hammat} \\
    P^+ & = & 2 \int_{- L}^{+ L} \!\! dx^- {\rm Tr}
    \ \bigl({\bf D}_- \Phi \bigr)^2
  \equiv {\pi\over  L} \hat K \ .
\label{mommat}
\end    {eqnarray}
The quantity $ \hat K$ is independent of the two dimensionful
parameters $ L $ and $ g $.
The Hamiltonian describes the interaction of two scalar matter currents
via an instantaneous gluon-like interaction
\cite{DKB93,BDK93}.
The instantaneous gluon is ``dressed'' by the zero mode of $ {\bf A}^+ $.
This zero mode, a color singlet object as shown in
Appendix \ref{sec:GribFock}, acts like a ``screening mass''.
Above all, $ 1/ {\bf D}_-^2 $ is never singular.
The system has what we will
call a set of pseudo-ground states generated by the zero mode
operator $ {\bf V}  $.
All these states have zero $ P^+$ but only one of them
is the true ground state having zero $ P^-$ upon
subtraction of the zero point energy. On all these
pseudo-vacua there will be a rich structure of matter states that
are allowed by Gauss' law.
For example, they include, but are not exhausted by the string states
discussed in \cite{DKB93,BDK93}.
\section{Quantization and Matter Currents for SU(2) Gauge Theory.}
%---+----1----+----2----+----3----+----4----+----5----+----6----+----7
The construction of the field $ \Phi $ is quite complicated
in $SU(N)$ for any $ N>2 $.
This is the primary stumbling block to formulating a large
$ N$ analysis in the presence of the zero mode of
$  {\bf A}^+$.
The extension to $SU(N)$ in the Chevalley basis
\cite{Dun92,DJP92} is easier and will be presented in a future work.
In the sequel we will thus proceed to analyze the model for $SU(2)$.
We will use a color helicity basis of the form
\begin {equation}
  \Phi = \tau ^3 \varphi_3 + \tau ^+ \varphi_+ + \tau ^- \varphi_-
\end    {equation}
for all field matrices.
This is explained in more detail in Appendix \ref{sec:notapp}.

%{\it The quantization of the gauge mode $ {\bf A}^+$.}
By gauge choice, the zero mode matrix $ {\bf V} $ is diagonal, thus
$  {\bf V}  = v \   \tau _3 \ . $
The component $ v \equiv v  (x^+)$ is a quantum
mechanical operator which we treat in the manner of
\cite{Man85}. We have previously encountered it in DLCQ in \cite{KPP94}
where, we showed that in the absence of dynamical quanta it
is the basis of a topological field theory isomorphic to equal-time
quantization.
The quantum $ v $ has a conjugate momentum
$  p \equiv \delta L / \delta  v  = 2 L \partial_+ v  $.
and satisfies the commutation relation
\begin {equation}
 \bigl[ v , p \bigr] =
 \bigl[ v , \ 2 L \partial_+ v \bigr] = i \ .
\end    {equation}
In the following it will be useful to invoke the dimensionless
combination
\begin {equation}
   z \equiv {g  v  L \over \pi} \ .
\end {equation}
Gribov copies then correspond to $ z \rightarrow  z + n_0$
for $ n_0$ some even integer. The odd integers are `copies' generated by
the group of centre conjugations of SU(2),
namely $Z_2$ symmetry \cite{Lus83}.
The finite interval $0 <  z  < 1$ is called the
{\it fundamental modular domain}, see for example \cite{vBa92}.
We emphasize the otherwise trivial fact that $ z $ is an
operator, better denoted $\hat{z}$.
In the subsequent analysis it is understood that
we work in a representation which diagonalizes that operator.
Thus $\hat{z} | z '\rangle =  z ' | z '\rangle$ and
$\langle  z ' | \hat{p} | z ''\rangle =
- i {\frac{\partial}{\partial z '}} \delta(z ' -  z '')$.
In the sequel we shall drop the delta functions, that is we shall
work in Schr\"odinger representation.

%{\it The quantization of the real scalar field $ \varphi_3$.}
The diagonal components of the hermitian color matrix $ \Phi $
are $ \varphi_3$.
The quantization, with the exception of the zero mode
$ {\hbox{\vbox{\ialign{#\crcr
    ${\,\scriptstyle \circ\,\,}$\crcr
   \noalign{\kern1pt\nointerlineskip}
    $\displaystyle{\varphi}_{3}$\crcr}}}}
=  a_0 / \sqrt{4 \pi} $, is canonical.
Any real-valued boson field
subject to periodic boundary conditions can be represented by
\begin {equation}
  \varphi_3 (x^+, x^-) = {a_0(x^+)\over \sqrt{4\pi}}
   + {1 \over \sqrt{4\pi}} \sum_{n =1}^{\infty} \Bigl(
  \  a_n (x^+)\  w_n \ {\rm e}^{-i n   {\pi\over L}   x^-}
  + \  a^\dagger_n (x^+)\  w_n \ {\rm e}^{+i n   {\pi\over L}   x^-}
  \Bigr) \ .
\end    {equation}
The factor $1/\sqrt{4\pi}$ is of course arbitrary but will turn
out to be convenient.
The notation $ a_n (x^+)$
should indicate that the creation and destruction operators
depend on the (light-cone) time.
The momentum field conjugate to $ \varphi_3$ is pure normal mode since
$ \pi^3 = \partial_- \varphi_3$.
The quantum commutation relation at equal $ x^+$ for the normal modes is
\begin {equation}
     \Bigl[
{\hbox{\vbox{\ialign{#\crcr
    ${\,\scriptstyle \circ\,\,}$\crcr
   \noalign{\kern1pt\nointerlineskip}
    $\displaystyle{\varphi}_{3}$\crcr}}}}
(x) , \pi^3(y) \Bigr]_{x^+=y^+} =
        {i\over2} \Bigl[ \delta  (x^- - y^-) - {1\over{2 L}} \Bigr] \ ,
\end    {equation}
where the last term ensures consistency for the commutator restricted to
normal mode fields \cite{KaP93}.
The Fock modes must consequently satisfy
$     \bigl[ a_n , a^\dagger_m \bigr] = \delta_n^m $
($n,m=1,\dots,\infty$)
and the coefficients must be $ w_n  = 1/ \sqrt{n} $.
The Kronecker $\delta_n^m $ is equivalent to
$\delta_{n , m}$.
The commutation relations of the
zero mode $ a_0 =  a_0^\dagger$ cannot be determined by any of
these relations: the mode obeys the `constraint equation' given below,
with a commutator
$[ a_0 , a_n ] \neq 0$ that must be solved for via the constraint.

The off-diagonal components of $ \Phi $
are complex valued operators with
$ \varphi_+ (x^+, x^-) = \varphi_-^\dagger (x^+, x^-) $.
With $ \pi^- = \bigl(\partial_- + i  g  v \bigr) \varphi_+$
as the momentum field conjugate to $ \varphi_-$, and
$ \pi^+ = \bigl(\partial_- - i  g  v \bigr) \varphi_-$
conjugate to $ \varphi_+$,
they obey the canonical commutation relations
\begin  {equation}
     \Bigl[ \varphi_-(x) , \pi^-(y) \Bigr]_{x^+=y^+} =
     \Bigl[ \varphi_+(x) , \pi^+(y) \Bigr]_{x^+=y^+} =
        {i\over2} \delta  (x^- - y^-)
\ . \label{fieldcomm} \end    {equation}
Following Franke \cite{FNP81,FNP82}, this is achieved by the
expansion over the momentum modes
\begin {equation}
   \varphi_-(x^+, x^-)
   = \sum_{n\in\ Z}
       {{\widetilde C}_n (x^+) \over {\sqrt{4\pi |n + z |}}}
     \ {\rm e}^{-in  {\pi\over L}   x^-} ,
     \quad {\rm{with}} \ %
       [ {\widetilde C}_n (x^+) , {\widetilde C}_{n^\prime}^\dagger (x^+) ]
   = \delta_n^{n^\prime} {\rm sgn} (n + z)
\ , \label{comcom} \end  {equation}
where the set of all integers is denoted by
$\ Z \equiv \{0,\pm 1, \pm 2, \dots, \pm \infty \} $,
as opposed to set of all half-integers
$ H \equiv \{\pm {1\over 2}, \pm {3\over 2}, \dots, \pm \infty \}$,
to be used below.
We want to go beyond Franke by explicitly introducing
particle and hole operators.
But this is met with difficulties,
because $ {\rm sgn} (n + z) $ is a very asymmetric function of $n$
for arbitrary values of $ z $. It is however always
possible to shift the summation index in Eq.(\ref{comcom}) by
$n = m - m_0$, with an arbitrary but given value of $m_0\equiv m_0(z)$.
Relabelling the operators $ \widetilde C_{m-m_0} \equiv  C_m$
by matter of convention gives identically
\begin {equation}
   \varphi_-(x)
   = {{\rm e}^{+ i m_0  {\pi\over L}   x^-} \over \sqrt{4\pi}}\
     \sum_{m}
       {{C}_m  (x^+) \over {\sqrt{| m + z  - m_0|}}}
     \ {\rm e}^{-i m   {\pi\over L}   x^-} ,
     \ {\rm{with}} \  [ C_m , C_{m^\prime}^\dagger ]
   = \delta_m^{m^\prime} {\rm sgn} (m + z  - m_0)
\ . \label{boscom} \end  {equation}
Now we need only define the shift constant $m_0(z)\in H $
in terms of
the `stair-function' ${\rm st}(z)$
\begin {equation}
       m_0(z) = {\rm st}(z) - {1\over 2} , \quad
     \zeta (z) = z - m_0(z) , \quad {\rm with}\ \ %
       {\rm st} (z) = \cases {[z] + 1 &for $z \ge 0$, \cr
                               [z]  &for $z  <  0$. \cr}
\label {staircase} \end   {equation}
Since $ {\rm st}(z) + {\rm st}(-z) = 1 $,
we can derive the fundamental relations:
\begin {equation}
  m_0 (z+1) = m_0 (z) + 1, \quad m_0  (-z)= -m_0  (z), \quad
\zeta  (z+1) = \zeta  (z), \quad \zeta (-z)= -\zeta (z)
 \ . \label {zsymm} \end   {equation}
Important is that $ - {1\over 2} < \zeta (z) < {1\over 2} $
for all values of $ z $
because this allows one to rewrite Eq.(\ref{boscom}) as
\begin {equation}
  \varphi_- (x) =
    {{\rm e}^{+ i m_0  {\pi\over L}   x^-} \over \sqrt{4\pi}}\ %
  \sum_{m={1\over 2}}^{\infty} \Bigl(
 \  b_m \  u_m \ {\rm e}^{-i  m   {\pi\over L}   x^-}
 + \  d^\dagger_m \  v_m \ {\rm e}^{+i  m   {\pi\over L}   x^-} \Bigr)
\ , \label{fockexp} \end    {equation}
{\it i.e.} in terms of particle and antiparticle operators,
$ b_m$ and $ d_m$, respectively.
The analogy with the (complex) Dirac spinor components \cite{EPB87}
is intentional, but the present particles obey boson
commutation relations
\begin {equation}
     \bigl[ b_n , b^\dagger_m \bigr] =
     \bigl[ d_n , d^\dagger_m \bigr] =
     \delta_{n}^{m} \ , \quad\  {\rm and} \quad
     \bigl[ b_n , d_m \bigr] =
     \bigl[ b_n , d^\dagger_m \bigr] = 0
\;. \label{focomrel} \end    {equation}
The real coefficients $ u_m (z) = 1/\sqrt{m + \zeta}$ and
$ v_m (z) = 1/\sqrt{m - \zeta}$ depend on $ z $ through $\zeta $.
Finally one notes that a large gauge transformation
$z\rightarrow z + 1$ produces only $m_0 \rightarrow m_0 + 1$
and thus only a change of the overall phase in Eq.(\ref{fockexp}).
Most importantly it does {\it not} change the particle-\-hole
assignment and thus the Fock vacuum defined with respect to
$b_m$ and $d_m$ is {\it invariant} under these transformations.

The Gauss law can be rewritten in terms of its explicit components,
namely
\begin {equation}
   -\partial_-^2  A_3 =  g  J^+_3, \qquad\quad
   -(\partial_- + i g   v)^2  A_+
   = g  J^+_+
\ , \label {gausscomp} \end   {equation}
and the hermitian conjugate of the latter with
$(J^+_+)^\dagger \equiv J^+_-$.
One would like to invert them to express $ A_3$ and
$ A_\pm$ in terms of the currents $J \equiv J _{\rm M}$,
which according to Eq.(\ref{eqofmot}) are defined as
\begin {equation}
  J^+_3 = {1\over i} \bigl(
           \varphi_+ \pi_- - \varphi_- \pi_+ \bigr)_s
 \ \  {\rm and}\quad
  J^+_+ = {1\over i} \bigl(
           \varphi_3 \pi_+ - \varphi_+ \pi_3 \bigr)_s
\ . \end    {equation}
The index $s$ indicates that noncommuting operators in this product
and in general must be symmetrized in order to preserve hermiticity.
Before the inversion of Eqs.(\ref{gausscomp})
one should investigate the zero mode structure of the currents.
An effective way to do so is to consider
their Discrete Fourier Transforms $\widetilde J$,
defined for convenience by
%---+----1----+----2----+----3----+----4----+----5----+----6----+----7
\begin {equation}
  J_3^+ (x^-) \equiv -{1\over 4 L} \sum_{k \in \ Z}
   e^{-i k {\pi \over  L} x^-} \ J_3^+ (k)
\ , \qquad {\rm and} \quad
  J_\pm^+ (x^-) \equiv -{1\over 4 L} \sum_{k \in \ H}
   e^{-i k {\pi \over  L} x^-} \ J_\pm^+ (k)
\ . \label{Fourier} \end    {equation}
One verifies that that
$ \bigl(J^+_3(k) \bigr)^\dagger =
        J^+_3(- k) $ and
$ \bigl(J^+_-(k) \bigr)^\dagger =
        J^+_+(- k) $.
In order to notationally disentangle $ a_0$ from the dynamic
modes we introduce composite
`charge operators' $ Q $ which are independent of $ a_0$
and the symmetrized remainders
$ B_k \equiv (a_0  b_k + b_k   a_0)/2 $ and
$ D_k \equiv (a_0  d_k + d_k   a_0)/2 $, {\it i.e.}\
\begin {equation}
   J^+_3(k) \equiv  Q_3 (k) , \quad
   J^+_+(k) \equiv  Q_+ (k)
     + {D_k \over  v_k}       , \quad\  {\rm and}\quad
     -J^+_-(k) \equiv  Q_+ (k)
     + {B_k \over  u_k}
\ . \label{defcharge}
\end    {equation}
The explicit expressions for the operators $ Q $ can be found
in Appendix \ref{sec:charges}.

Because of the boundary conditions,
the first of the Gauss equations (\ref{gausscomp}) can be solved
only if the zero mode $< J^{+}_{3} >_o \equiv  Q_0 $
on the r.h.s vanishes.
This cannot be satisfied as an operator, but must be used to
select out physical states, {\it i.e.}\
$     Q_0 \vert {\rm {phys}} \ \rangle \ \equiv 0 $.
In second-quantized form this gives
\begin {equation}
      Q_0 \vert {\rm {phys}} \rangle =
   \sum_{m = {1\over 2}}^{\infty}
   \Bigl(b^\dagger_m   b_m - d^\dagger_m   d_m \Bigr)
    \vert {\rm {phys}} \rangle = 0 \ .
\label{gaussop}
\end    {equation}
It is thus simple to find states satisfying this: they
must have the same total number of `` b '' and `` d '' particles.
The resemblance to the electric-charge neutrality condition
is because the residual global gauge symmetry we are factoring out
of the Hilbert space is, as mentioned earlier, Abelian.

To complete the specification of the
Hilbert space we give the momentum operator in second quantized form.
Evaluating the trivial integrals in Eq.(\ref{mommat}),
one gets the dimensionless operator
\begin  {equation}
 \hat K =
    \sum_{n =1}^\infty
      {n}\  a^\dagger_n   a_n
    + \sum_{m = {1\over2}}^\infty \Bigl[
      (m+\zeta)\  b^\dagger_m   b_m + (m-\zeta)\  d^\dagger_m  d_m  \Bigr]
\,. \label{secqmom} \end    {equation}
%---+----1----+----2----+----3----+----4----+----5----+----6----+----7

We finally turn to the constraint equation.
Taking the zero mode of the matter field equation (\ref{phieqm})
explicitly yields three equations.
Two of them give true dynamical equations for the zero modes
${\hbox{\vbox{\ialign{#\crcr
    ${\,\scriptstyle \circ\,\,}$\crcr
   \noalign{\kern1pt\nointerlineskip}
    $\displaystyle{\varphi}_{\pm}$\crcr}}}}
$, which for reasons given above we do not
wish to solve explicitly so we do not give them here.
The third equation, with
$  c \equiv 8\pi {\rm Tr} \
  <   \tau ^-  {\bf D}^\alpha  {\bf D}_\alpha \Phi   >_o / g^2$,
becomes after some algebra
%---+----1----+----2----+----3----+----4----+----5----+----6----+----7
\begin {equation}
    c  = \Bigl\langle \
       \varphi_+ {1\over (\partial_- - i  g   v)} J^+_-
     - \varphi_- {1\over (\partial_- + i  g   v)} J^+_+
     \ \Bigr\rangle_{0,s}   = 0 \ ,
\end    {equation}
and is the constraint equation.
Insertion of the above yields
%---+----1----+----2----+----3----+----4----+----5----+----6----+----7
\begin {eqnarray}
  \sum_{k = {1\over 2}}^{\infty} \Bigl[
     u_k^2 \Bigl(B^\dagger_k  b_k  + B_k   b^\dagger_k \Bigr)_s
  & + &
     v_k^2 \Bigl(D^\dagger_k  d_k  + D_k   d^\dagger_k \Bigr)_s
\Bigr] \nonumber \\
=   - \sum_{k = {1\over 2}}^{\infty} \Bigl[
     u_k^3 \Bigl(Q^\dagger_-(k) b_k + b^\dagger_k  Q_-(k)\Bigr)
  & + &
     v_k^3 \Bigl(Q^\dagger_+(k) d_k + d^\dagger_k  Q_+(k)\Bigr)
\Bigr] \ . \label{fullconstraint} \end   {eqnarray}
This is the most compact expression for the constraint.
It is clearly linear in $ a_0$ and therefore quite different in structure
from the constraint equation of $\phi^4_{1+1}$.
It is not clear then how it could give
rise to spontaneous symmetry breaking in the scenario of
\cite{Rob93,Hei91,Hei92a,Hei92b,BPV93,PiV94a,HPV94}.
Despite its linearity, Eq.(\ref{fullconstraint})
is still complicated to solve as a quantum operator constraint.
At this point, we therefore isolate this part of the
overall problem and return to it in a future treatment.
For example, under active consideration now is a
solution via Fock space truncation methods as employed in
\cite{BPV93,PiV94a,HPV94}.

We end this section by summarizing the zero mode `zoo' of this theory.
The complex field $\varphi_+$ actually has no true zero mode
in the sense of vanishing eigenvalue of $ P^+$.
Secondly, there is the topological zero mode $ z $
which must be treated by diagonalizing its Hamiltonian.
Third, there is the constrained zero mode, $ a_0$.
It must be solved at the level of the constraint equation
(\ref{fullconstraint})  and its presence in
the Hamiltonian eliminated in favour of the true degrees of freedom.
%---+----1----+----2----+----3----+----4----+----5----+----6----+----7----+
%
\section{The Hamiltonian for Pure SU(2) Gauge Theory.}
%---+----1----+----2----+----3----+----4----+----5----+----6----+----7
The front form Hamiltonian, Eq.(\ref{hammat}), rewritten in
terms of the components becomes
\begin {equation}
    P^- =  L  (\partial_+  v)^2
    + {1 \over2} \int_{- L}^{+ L} \!\!\!\! dx^- \Bigl[
    \partial_-  A_3 \partial_-  A_3
    + (\partial_- +i g  v)  A_+
      (\partial_- -i g  v)  A_-
    + (\partial_- -i g  v)  A_-
      (\partial_- +i g  v)  A_+             \Bigr] \ .
\end    {equation}
Note that the zero mode of $ A_3$ does not occur since
this field always appears here acted upon by a space derivative.
Because of periodic boundary conditions one can integrate by parts
and eliminate $ A $ in favour of
$ J \equiv  J_{\rm M}^+$ using Eq.(\ref{gausscomp}), {\it i.e.}\
\begin {equation}
    P^- = {p^2 \over 4 L} - {g^2 \over 2}
 \int_{- L}^{+ L} \!\!\!\! dx^- \Bigl[
{\hbox{\vbox{\ialign{#\crcr
    ${\,\scriptstyle n\,\,}$\crcr
   \noalign{\kern1pt\nointerlineskip}
    $\displaystyle{J}_{3}$\crcr}}}}
{1\over \partial_-^2}
{\hbox{\vbox{\ialign{#\crcr
    ${\,\scriptstyle n\,\,}$\crcr
   \noalign{\kern1pt\nointerlineskip}
    $\displaystyle{J}_{3}$\crcr}}}}
  + J_+ {1\over (\partial_- -i g  v)^2} J_-
  + J_- {1\over (\partial_- +i g  v)^2} J_+
                                        \Bigr] \ .
\end    {equation}
Notice that the operator has no ill-defined singularities.
This expression is best written in discrete Fourier space.
We can factor out all dimensionful parameters by
$ P^- =  L \Bigl(g  / 4\pi\Bigr)^2  H $,
which defines the dimensionless Hamiltonian $ H $.
The Discrete Fourier Transforms Eq.(\ref{Fourier}) of the
currents can be used and the subsequent momentum sums expressed
over positive values of $ k $ by
$ \widetilde J_3(- k) = \widetilde J^\dagger_3 (k) $,
$ \widetilde J_+(- k) = \widetilde J^\dagger_- (k) $, and
$ \widetilde J_-(- k) = \widetilde J^\dagger_+ (k) $.
In terms of the coefficients $ u $ and $ v $
the result is
\begin  {eqnarray}
 H  =
 - 4{d^2 \over d z^2}
 + \sum_{k =1}^{\infty}
     w_k^4 \bigl(\widetilde J_3^\dagger (k) \widetilde J_3 (k)
         + \widetilde J_3 (k) \widetilde J_3^\dagger (k) \bigr)
 & + & \sum_{k ={1\over2}}^{\infty}
     v_k^4 \bigl(\widetilde J_+^\dagger (k) \widetilde J_+ (k)
         + \widetilde J_+ (k) \widetilde J_+^\dagger (k) \bigr)
\nonumber \\
 & + & \sum_{k ={1\over2}}^{\infty}
     u_k^4 \bigl(\widetilde J_-^\dagger (k) \widetilde J_- (k)
         + \widetilde J_- (k) \widetilde J_-^\dagger (k) \bigr)
  \ .
\label{hamall}
\end   {eqnarray}
This result resembles in many respects the structure found in treatments
of gauge theory on a `cylinder' in standard instant form Hamiltonian
quantization \cite{LaS92}.
Insofar as \cite{DKB93,BDK93} omit zero modes, it disagrees with
their expression for the Hamiltonian.
We now look in more detail at the separate contributions
to this expression from, respectively, the gauge mode $ z $,
the Fock operators and the constrained zero mode.
In particular, in the next section we shall be
especially interested in the VEV of the Hamiltonian. This
will guide our present analysis.

First we consider the pieces which do not involve the constrained
zero mode $ a_0$. This is simply achieved by replacing, in the
above Hamiltonian, the currents $\widetilde J $ with the operators $ Q $ via
the
definition (\ref{defcharge}).
Inspecting the expressions for $ Q $
given in Appendix \ref{sec:charges} one observes they all
annihilate the Fock vacuum $\vert 0 \rangle$,
while $ Q^\dagger \vert 0 \rangle \neq 0$.
So the full Hamiltonian has a VEV,
henceforward denoted by $ V_0 (z)$.
Since it depends on the {\it quantum operator} $ z $, $ V_0 $ cannot be
removed by the usual trivial vacuum renormalization.
Rather the VEV plays the role of a `potential energy' in
what we call the `gauge part' of the Hamiltonian
which also includes the kinetic term for $ z $, namely
\begin {equation}
     H_{Gauge} \equiv
- 4 {d^2 \over d z^2} + \langle 0| H |0\rangle \equiv
- 4 {d^2 \over d z^2} +  V_0 (z) \ .
\label{gaugepot}
\end   {equation}
We extract the expression for $ V_0 (z)$ by bringing
the Hamiltonian to normal ordered form.
Commuting the operators $ Q^\dagger$ with $ Q $
in the substitute of Eq.(\ref{hamall}) leaves us
with the contribution $ H_{Fock}^{(1)} $
to the Fock space part of the Hamiltonian
$ H_{Fock} = H_{Fock}^{(1)} +   H_{Fock}^{(2)}$, {\it i.e.}\ %
\begin {equation}
  H_{Fock}^{(1)} \equiv
   2\sum_{k =1}^{\infty}
     w_k^4\  Q_3^\dagger (k)  Q_3 (k)
 + 2\sum_{k ={1\over2}}^{\infty}
     v_k^4\  Q_+^\dagger (k)  Q_+ (k)
 + 2\sum_{k ={1\over2}}^{\infty}
     u_k^4\  Q_-^\dagger (k)  Q_- (k)
  \ .
\end   {equation}
The VEV of the left-over commutator term generates the gauge mode potential
\begin {equation}
      V_0 \equiv
   \sum_{k =1}^{\infty}
      w_k^4\ \langle 0 \vert \Bigl[ Q_3 (k) ,
                        Q_3^\dagger (k) \Bigr] \vert 0\rangle
   + \sum_{k ={1\over2}}^{\infty}
      v_k^4\ \langle 0 \vert \Bigl[ Q_+ (k) ,
                        Q_+^\dagger (k) \Bigr] \vert 0\rangle
   + \sum_{k ={1\over2}}^{\infty}
      u_k^4\ \langle 0 \vert \Bigl[ Q_- (k) ,
                        Q_-^\dagger (k) \Bigr] \vert 0\rangle
     .
\label{gaugepoteq}
\end   {equation}
This is analyzed in detail in Appendix \ref{sec:gaugepot} and
represented in Fig.1.
Subtracting Eq.(\ref{gaugepoteq}) from the commutator expression gives the
other contribution to the Fock-space Hamiltonian:
\begin {equation}
      H_{Fock}^{(2)} \equiv
     \sum_{k =1}^{\infty}
        w_k^4\ \Bigl[ Q_3 (k) , Q_3^\dagger (k) \Bigr]
     + \sum_{k ={1\over2}}^{\infty}
        v_k^4\ \Bigl[ Q_+ (k) , Q_+^\dagger (k) \Bigr]
     + \sum_{k ={1\over2}}^{\infty}
        u_k^4\ \Bigl[ Q_- (k) , Q_-^\dagger (k) \Bigr]
    - V_0 (z)
   \ .
\end   {equation}
Thus far it would appear to suffice to
consider the $ Q $ operators and $ z $ for the determination
of the spectrum.

This changes if one addresses the constrained mode $ a_0$
as the constraint cannot be expressed purely in terms of $ Q  $
operators. From Eq.(\ref{hamall}) and (\ref{defcharge}),
one finds $ a_0$ makes its appearance in the Hamiltonian
both linearly and quadratically
which we separate into $ H_{Constr} = H_{Constr}^{(1)} +  H_{Constr}^{(2)} $.
The term linear in $ a_0$ is
\begin {equation}
   H_{Constr}^{(1)} =
   2 \sum_{k ={1\over2}}^{\infty} \Bigl[\ %
      u_k^3\ \Bigl(B^\dagger_k   Q_- (k) +
                      Q_-^\dagger(k) B_k \Bigr)_s
  + v_k^3\ \Bigl(D^\dagger_k   Q_+ (k) +
                      Q_+^\dagger(k) D_k \Bigr)_s \ \Bigr]
\ . \end   {equation}
The term quadratic in $ a_0$ is
\begin {equation}
    H_{Constr}^{(2)} = \sum_{k ={1\over2}}^{\infty} \Bigl[\ %
      u_k^2\ \bigl(B_k B_k^\dagger + B_k^\dagger B_k \bigr)
  + v_k^2\ \bigl(D_k D_k^\dagger + D_k^\dagger D_k \bigr)
\ \Bigr] \ . \end   {equation}
$ H_{Constr} $ could have a non-zero VEV that could contribute to
the potential of the gauge mode.
There are two possibilities where
such nonvanishing contributions could arise. The first is when
an $ a_0$ appears either to the extreme left or extreme
right in $ H_{Constr} $. However, preliminary studies of the structure
of the constraint suggest that no non-zero VEV of $ a_0$ or
$(a_0)^2$ can arise.
Thus, the role played by this mode is quite different from
that of its counterpart in the $(\phi^4)_{1+1}$ theory.
The second possibility of non-zero contribution to the
gauge potential is when the zero
mode lies between an annihilation operator on the left
and a creation operator on the right, such as
$ \langle 0| b_k   a_0  a_0  b^\dagger_k |0 \rangle$.
A nonzero VEV could potentially arise from such contributions
if $[ a_0, b_k ]$ and $[ a_0, b^\dagger_k ]$ are non-zero.

This completes the analysis of the light-cone Hamiltonian.
In summary though, we have reexpressed
the Hamiltonian as a sum of three contributions, {\it i.e.}\
\begin {equation}
    H = H_{Gauge} + H_{Fock} + H_{Constr}
\ . \label{hamsum} \end   {equation}
In the next section we will see how far analytic methods
can take us in diagonalizing parts of the Hamiltonian.

Finally, we should emphasize that we have refrained thus far
from {\it ad hoc} approximations to define a consistent Hamiltonian
and Hilbert space.
As we are unable to give here exact analytical solutions
we consider their numerical simulation as a challenge for the
future. In the sequel we therefore will continue with several
simplifying assumptions which will allow us to solve part of
the Hamiltonian analytically. The most drastic among them
is the omission of the constraint part.
We thus will concentrate efforts in the following on solving
approximately $ H_{Gauge} $ and $ H_{Fock} $  subject to
the omission of $ H_{Constr} $.
%---+----1----+----2----+----3----+----4----+----5----+----6----+----7

\section{Approximate Solutions in the Gauge Sector}
The potential energy $ V_0 (z)$ which appears in $ H_{Gauge} $ via
Eq.(\ref{gaugepot}), {\it i.e.} \
$ H_{Gauge} \equiv - 4 {d^2 / d z^2} +  V_0 (z)$,
is what we shall loosely refer to as the `gauge potential'.
As analyzed in Appendix \ref{sec:gaugepot} it is {\it invariant}
under large gauge transformations.
One therefore can restrict oneself to calculate it only for
the fundamental modular domain $0 <  z  < 1$.
In Fig.1 it is displayed for the two cases explained below.
In either case, because of the singular behaviour at $ z  = 0$ and
$ z  = 1$, there will be a discrete spectrum of
excitation energies in this potential. These will be labelled by
a quantum number $N$.
Generalizing the representation in \cite{KPP94} we use a wavefunctions
$\Psi_{N}(z) \equiv \langle  z |N\rangle$
so that we address ourselves to solving the Schr\"odinger equation
\begin  {equation}
  H_{Gauge} \Psi_{N}(z) = \widetilde E_{N} \Psi_{N}(z)
\ . \end    {equation}
The symbol $E_N$ will be reserved for the vacuum normalized
eigen-energy, namely $\widetilde E_{N} - \widetilde E_{0}$.

In the fundamental modular domain the potential
is symmetric about $ z  = {1\over2}$. As mentioned, it is singular
at $ z  = 0$ and $1$.
It has a minimum at $ z  = {1\over2}$, however the value of the function
there is actually divergent.
Moreover its curvature diverges at that point,
that is $ V_0 ''({1\over2}) \propto \omega_0^2 \ln{s}$,
$s \rightarrow \infty$, where
$s$ is the value of a dimensionless regulator truncating sums.
The precise meaning of this divergent behaviour, namely whether
it is physical or formal, is unclear to us.
Lacking a definite answer,
we pursue below the two alternative
scenarios, labelled (a) and (b).

\noindent
{\underline{Scenario (a) -- Cutoff Independent Approach:}}
Let us take the point of view that the divergent $\ln{s}$ behaviour of
the potential should be cancelled by a counterterm in the Lagrangian.
In fact we do not know what this counterterm should be
and the answer may come from the, as yet unavailable, solution
to the constrained zero mode $a_0$.
We therefore proceed by simply numerically subtracting
the function $\omega_0^2 \ln{s}\ (z  - {1\over2})^2 $
from the explicit expression
in Appendix \ref{sec:gaugepot}.
We plot the so-obtained function in Fig.1. Upon inspection,
one notes the very flat base in the vicinity of $ z =1/2$.
A good approximation might therefore be the infinite square well.
Restoring units, the problem is expressed by
$ -  L  (g /2 \pi)^2 (d^2/dz^2) \Psi_{N}
  = \widetilde E_{N} \Psi_{N}$.
After a vacuum energy subtraction, the eigen-energy {\it densities}
$\epsilon_{N} \equiv E_{N}/ (2  L)$
are $ g^2 (n^2 - 1)/8$.
We thus recover the spectrum for SU(2) pure zero mode glue
in 1+1 dimensions as obtained in \cite{Het93a,Het93b},
and verified by us on the light-cone in \cite{KPP94}.
Wavefunctions respecting boundary conditions are
then ${\rm{sin}}(N \pi  z)$ up to normalization.
As for subtleties concerning this choice over the cosine
we refer the reader to \cite{KPP94}.
Thus we have succeeded in diagonalizing
the gauge part of $ P^-$. As the vacuum part of $P^+$ has no
dependence on $z$ these wavefunctions are also eigenfunctions of
$P^+$ with momentum zero.

\noindent
{\underline{Scenario (b) -- Cutoff-Dependent Approach:}}
For reasonably large values of the cutoff $s$, the $\ln{s}$ term
dominates, as mentioned, over the $s$-independent part
and thus here the latter can be omitted.
As discussed in Appendix \ref{sec:gaugepot}, with the cutoff
finite but large the gauge potential can be approximated by
\begin  {equation}
   V_0 (z) =  V_0 ({\textstyle{1\over2}})
           + 4\omega^2 (z -{\textstyle{1\over2}})^2\ ,
 \quad\ {\rm with}\quad  \omega^2 = \omega^2_0 \ln{s} \ .
 \label{harmonic} \end    {equation}
The numerical value of $\omega_0^2$ is given in
Appendix \ref{sec:gaugepot},
while ${V_0}({1\over2})$ is an unspecified constant.
The eigenvalues and eigenfunctions
are now those of the harmonic oscillator,
\begin  {equation}
 \widetilde E_{N} =  V_0 ({\textstyle{1\over2}}) +  4\omega (2N+1) \ ,
 \quad\  {\rm and} \quad
  \Psi_{N} (z) = {\cal N}_N
     {\rm H}_N \Bigl(\sqrt{\omega} (z -{\textstyle{1\over2}}) \Bigr)
  \,{\rm e}^{-{\omega \over 2} (z - {1\over 2})^2} \ ,
\end    {equation}
respectively, where the $ {\rm H}_N$ are the Hermite polynomials
of order $N$, and where the ${\cal N}_N $
normalize all eigenfunctions to unity.
The state with lowest energy has $N=0$, by inspection,
and has to be identified with the true `vacuum'.
The vacuum-renormalized eigenvalues and
lowest energy eigenfunctions are thus
\begin  {eqnarray}
   E_{N} & \equiv & \widetilde E_{N} - \widetilde E_{0} =
   8 N \omega_0 \sqrt{\ln {s}} \ ,
  \quad\  {\rm and} \\
  \Psi_{0} (z)
    & = & \Bigl({\omega \over \pi} \Bigr)^{{1\over 4}}
    \,{\rm e}^{-{\omega \over 2} (z  - {1\over 2})^2}
   \sim \sqrt{\delta (z -{\textstyle{1\over2}})} \ .
\end    {eqnarray}
In the limit $s\rightarrow\infty$ the eigenfunctions
degenerate into, roughly speaking, $\delta$-`functions',
namely the wavefunctions strongly localize about $ z  = {1\over2}$.
This extremely sharp peaking implies that the {\it operator} $ z $
can be replaced by the {\it number} ${1\over2}$ everywhere but in
the gauge Hamiltonian, a significant simplification.
It is what makes this scenario radically different from the former.
As a consequence,
the gauge excitations $\Psi_{N}(z)$, the ``gaugeons'',
have for finite but sufficiently large $s$ a
finite and non-degenerate (light-cone) energy but strictly zero mass.

\section{Approximate Solutions to the Gauge plus Fock Sector}
%---+----1----+----2----+----3----+----4----+----5----+----6----+----7
In the remainder we address ourselves to finding eigenstates to
$ H_{Gauge} + H_{Fock}  $.
In the absence of a complete picture for the renormalization in
scenario (a), we implement here the conclusions of scenario (b)
for the gauge mode:
we substitute $ z  = {1\over2}$ where appropriate.
The Hilbert space will be spanned by the product states of gauge
eigenfunctions $\Psi_{N}(z)$ and `Fock states'. The product
states $\Psi_{N}(z) |0\rangle_{\rm{Fock}}$ are what we
loosely call {\it pseudovacua}. The true vacuum is
$|0\rangle = \Psi_{0} |0\rangle_{\rm{Fock}}
\sim \sqrt{\delta(z  - {1\over2})} |0\rangle_{\rm{Fock}}$.
On top of it we now build Fock space excitations.

Let us consider three types of two-particle Fock-space excitations,
the toy states
\begin  {equation}
 | {k ;3} > \rangle \equiv   Q_3^\dagger (k)
| {0} > , \quad
 | {k ;+} > \equiv  Q_+^\dagger (k) | {0} > ,\ \quad {\rm{and}}\quad
 | {k ;-} > \equiv  Q_-^\dagger (k) | {0} > , \quad
\end    {equation}
to which we shall refer collectively as $| k ; a \rangle$.
We mention already here that only
one of these, $| k ;3 \rangle$, satisfies Gauss' law.
Using the approximate delta-function behaviour of the gauge
wavefunctions, the toy states are seen to be eigenstates of $ P^+$ {\it i.e.}
\begin   {equation}
  \hat K | {k ;3} > =   k | {k ;3} > ,\quad
  \hat K | {k ;+} > =   k | {k ;+} > ,\quad {\rm{and}} \quad
  \hat K | {k ;-} > =   k | {k ;-} >
\ .\end    {equation}
Are they also eigenstates to the full $ P^-$?

In general, the commutators of charge operators $ Q_a (k) $
acting on the vacuum can be separated into c-number and operator parts
\begin  {equation}
 \Bigl [ Q_a (p) , Q_b^\dagger (k) \Bigr] | {0} >
   = (\delta_{ab}  S_a(k) \delta^k_p
   + {\cal O}_{ab}(p , k)) | {0} >
\ , \label {ModelComm} \end    {equation}
with the c-number coefficients $ S_a$ defined in
Appendix \ref{sec:gaugepot}.
The operator part ${\cal O}_{ab}(p , k)$ is complicated
to write down in full and is generally non-zero.
However, for $ p = k $, ${\cal O}_{ab} = 0$.
Taking this as a hint, we shall {\it assume}
that the effects of ${\cal O}_{ab}$ are in some sense ``small''
and set the entire operator to zero by hand.

With these simplifications now, the commutator $ H_{Fock}^{(2)}$
vanishes and all three toy states $| k ; a\rangle$ become
eigenstates of $ H_{Fock}^{(1)} $, like for example
\begin  {eqnarray}
 H_{Fock}^{(1)} | {k ;3} > & = &
   2\sum_{p ={1\over2}}^{\infty}
     v_p^4\  Q_+^\dagger (p)
  \Bigl[ Q_+ (p) , Q_3 (k)^\dagger \Bigr] | {0} >
 + 2\sum_{p =1}^{\infty}
     w_p^4\  Q_3^\dagger (p)
  \Bigl[ Q_3 (p) , Q_3 (k)^\dagger \Bigr] | {0} >
\nonumber \\
 & + & 2\sum_{p ={1\over2}}^{\infty}
     u_p^4\  Q_-^\dagger (p)
  \Bigl[ Q_- (p) , Q_3 (k)^\dagger \Bigr] | {0} >
  \quad \simeq \quad
    4 \ln  k | {k ;3} >
\; ,
\end{eqnarray}
for sufficiently large values of $ k $,
according to Eqs.(\ref{cc3func}).
Recall that the continuum limit is reached \cite{PaB85a,PaB85b}
by the limit $ k \rightarrow \infty$.
The combined action of the energy and momentum operators
$ \hat H \equiv \hat K (H_{Gauge} + H_{Fock}) $ becomes thus
$ \hat H | {k ;a} > = 4 \ln {k} | {k ;a} > $.
The action of the mass-squared operator $ P^+ P^-$
on the toy states gives finally,
after restoring units according to Eq.(\ref{transcale}),
\begin  {eqnarray}
  P^+  P^- | {k ;a;N} > & = &
{g^2 \over {8\pi  L_{\!\bot}^2}}
 \Bigl(\ln{k}
 + 2  k  N \omega_0 \sqrt{\ln{s}} \Bigr)
 | {k ;a;N} > \ , \quad
\\
 \ {\rm with} \qquad | {k ;a;N} > & = &
   Q_3^\dagger (k) \Psi_{N}(z)
 | {0} >_{{\rm Fock}} \ ,
\label {p+p-state}
\end    {eqnarray}
independent of the longitudinal interval length $ L $.
This then is an approximate mass spectrum of the
model in the two-particle sector with all cutoffs large but finite.
How does it behave
as cutoffs are removed? We take the necessary limits
as follows. (1) There is no meaning to the
transversal continuum limit in the present model so we consider
$ L_{\!\bot} $ as arbitrary but fixed. (2) Since it is meaningful
to consider physics in a finite volume or interval, $s$
should be taken to its physical limit before $ L $. This
removes from the spectrum all the ``gaugeon'' excitations $N$.
(3) As mentioned, the longitudinal
continuum limit is defined by
$ k \rightarrow \infty, L \rightarrow \infty$, but $ p^+ = \pi  k  /  L $
fixed. Since the longitudinal length does not appear one has to
take the isolated limit $ k \rightarrow \infty$.
Thus the degenerate triplet of states with $N = 0$
also diverge in the continuum limit
and do not survive in the bare spectrum.

\section{Discussion and Perspectives}
%---+----1----+----2----+----3----+----4----+----5----+----6----+----7
Let us briefly restate the approach we took here.
Beginning with SU(2) gauge theory in (2+1) dimensions in the
front form we suppressed transverse momenta
of the gluons and obtained a (1+1) dimensional gauge theory
coupled to adjoint scalar matter. Gauge fixing of this
theory revealed for the content many dynamical
normal modes of the scalar field, a topological gauge zero mode,
and a constrained zero mode.
The constrained mode satisfies a linear but nonetheless
complicated operator constraint.
The gauge-fixing involved a space-time independent
color rotation that reduced the remnant Gauss law to be implemented
to an Abelian global symmetry generator.
We succeeded in specifying the space of states which would
be annihilated by the Gauss operator, namely that of color singlet
states built from the Fock or parton operators.
Not performing the
gauge rotation would not even permit one to easily solve
Gauss' law. We succeeded in diagonalizing both $ P^+$
and $ P^-$ in the gauge mode sector in two separate approaches
which respectively involved keeping or removing an ultraviolet cutoff in the
calculation. The cutoff independent approach lead to gauge mode
wavefunctions that were unlocalized.
With the cutoff, the
solution of the gauge mode problem in the Fock space sector
reduced to substituting in the Fock sector
the minimum value of its potential by the value $\zeta =0$.
We approximately and analytically solved for the invariant mass
of three composite states, one of which satisfied Gauss' law.
Even for this one, the energy diverged in the
continuum limit.

We now interpret the meaning of this result.
It might be that some aspect of the
nontrivial renormalization required in full (2+1) dimensions
manifests itself even in this (1+1)-dimensional sub-regime.
While the theory is superrenormalizable by virtue of dimensionality
the structure of the (finite) number of divergences is not of the usual
two-dimensional QCD type but reflect some substructure of the higher
dimensional theory and its renormalization and scaling properties.
Recalling that every Lagrangian field theory has
an open scale, only {\it mass ratios} can be meaningful quantities.
If our toy states reflect correctly the behaviour of
the lowest energy singlet state, namely running like
${{g^2} \over {4\pi}} \ln  k $, then renormalization of the
spectrum is achieved by `renormalizing the coupling constant'. This would
then read explicitly
\begin{equation}
 g^2 =  g^2_{\rm{phys}}/\ln  k
\end{equation}
leading to a lowest excitation of mass `1' in arbitrary units.

We now return to the question of the diagonalization of the
gauge part of the Hamiltonian in scenario (a).
It leads to a more complicated treatment of the $ z $-mode
when including the Fock space excitations.
There is nothing in principle hindering a full solution
of this but we leave it for future work.
What remains to be understood is what type of counterterm
could remove this divergence. It is possible
that the omitted constrained mode contributions assist in the
renormalization of the potential. We are presently exploring
this in the context of Fock space truncation approximations
\cite{BPV93,PiV94a,HPV94} for solving the constraint.
That the constrained zero mode can provide
renormalization counterterms is not without precedent, for example this
has already been seen in perturbative QED in \cite{KaR94}.
On the other hand, the picture emerging in scenario (b),
of a special role for
the value $\zeta = 0$, has also been seen via point-splitting
regularization of Gauss' law and the momentum operator. This
will be reported elsewhere \cite{PKP95}.

More can be done analytically:
evidently some nontrivial linear combination of the $ Q_a(k)$
operators could build a color singlet for which an approximate
eigenvalue might be obtainable.
Nevertheless, the treatment of the first four sections has
prepared the way for treating the theory with the full power
of {\it standard} DLCQ numerical techniques, now including zero modes.
Such numerical work is underway.

Going beyond the present theory would mean addressing dimensionally
reduced QCD(3+1) where all the features discussed here will
continue.
The hope that DLCQ can allow us to understand QCD in an intuitively
simple way but with its full richness remains
undiminished.

\section{Acknowledgement}
%---+----1----+----2----+----3----+----4----+----5----+----6----+----7
The authors acknowledge J. Hetrick and L. Hollenberg for
advice.
We especially thank B. van de Sande for assistance
in preparing Fig.1.
(SSP) would like to thank the Max-Planck Institut f\"ur
Kernphysik for its hospitality during his visit. Both
(ACK) and (HCP) thank the Department of Physics, The Ohio-State
University, for its support and hospitality during various
visits for the purpose of this work.
(ACK) was supported by the DFG under contract DFG-Gz: Pa 450/1-2.
Travel support was provided in part by a NATO Collaborative Grant.
%---+----1----+----2----+----3----+----4----+----5----+----6----+----7

 \begin {appendix}
 \newpage
\section{Notation and Conventions}
\label{sec:notapp}
%---+----1----+----2----+----3----+----4----+----5----+----6----+----7

\noindent
{\underline {Equations and Constants of Motion for QCD.}}
In quantum chromodynamics the gauge fields are traceless
hermitean $3\times 3$ {\it matrices} $ {\bf A}^\mu $.
More generally for SU(N), they are $N \times N$ matrices
parametrized in terms of `color vector
potentials' $ A^\mu_a$, {\it i.e.}\ $ {\bf A}^\mu \equiv T^a A^\mu_a$.
The glue index $a$ (or $r,s,t$) is implicitely summed with
no attention paid to the lowering or raising
and runs from $1$ to $N^2-1$.
The quark field $ \Psi $ is a color triplet spinor,
{\it i.e.}\ $ \Psi_{\alpha,c}\,$, but the Dirac index $\alpha$
and the color index $c = 1,\dots,N$ are usually suppressed.
The color matrices $ T^a_{cc^\prime}$ obey
$ \bigl[ T^a, T^b\bigr]_{cc^\prime}
  = i f^{abr}  T_{cc^\prime}^r $
and ${\rm Tr} \bigl(T^a  T^b \bigr) = {1\over2}\delta_{ab}$.
They are related to the Gell-Mann
matrices $\lambda^a$ by $ T^a={1\over2} \lambda^a$.
For SU(2) the  $\lambda^a$ are the Pauli matrices $\sigma^a$,
and for SU(3) one has {\it e.g.}\ %
%---+----1----+----2----+----3----+----4----+----5----+----6----+----7
\begin {equation}
  ({\bf A}^\mu)_{cc^\prime} = {1\over2} \pmatrix{
   {1\over\sqrt3} A^\mu_8+ A^\mu_3
  & A^\mu_1-i A^\mu_2
  & A^\mu_4-i A^\mu_5 \cr
    A^\mu_1+i A^\mu_2
  &{1\over\sqrt3} A^\mu_8- A^\mu_3
  & A^\mu_6-i A^\mu_7 \cr
    A^\mu_4+i A^\mu_5
  & A^\mu_6+i A^\mu_7
  &-{2\over\sqrt3} A^\mu_8 \cr} \ .
\end {equation}
The Lagrangian density for QCD can thus be written in two
equivalent conventions
\begin {eqnarray}
    {\cal L}  = - {1\over2}
    {\rm Tr} \bigl({\bf F}^{\mu\nu} {\bf F}_{\mu \nu} \bigr)
 + {1\over2} \bigl[ \overline \Psi \bigl(i\gamma^\mu  {\cal D} m_\mu
   -  m \bigr)
  \Psi   + \ {\rm h.c.} \bigr] , \ {\rm with}\
    {\bf F}^{\mu\nu} \equiv \partial^\mu  {\bf A}^\nu
   - \partial^\nu  {\bf A}^\mu
   + i g \bigl[ {\bf A}^\mu, {\bf A}^\nu\bigr] , \\
   {\cal L}  = - {1\over4}  F^{\mu\nu}_a  F_{\mu \nu}^a
  + {1\over2} \bigl[ \overline \Psi \bigl(i\gamma^\mu
   {\cal D} m_\mu -  m \bigr)
 \Psi   + \ {\rm h.c.} \bigr] , \ {\rm with}\
   F^{\mu\nu}_a \equiv \partial^\mu  A^\nu_a
  - \partial^\nu  A^\mu_a
 - g  f^{ars}  A^\mu_r  A^\nu_s  .
\end {eqnarray}
The covariant derivative {\it matrix} is
$ \bigl({\cal D} ^\mu\bigr)_{cc^\prime}
 \equiv \delta_{cc^\prime} \partial^\mu
  + i g \bigl({\bf A}^\mu\bigr)_{cc^\prime} $.
The variational derivatives are
\begin {equation} {\delta  {\cal L} \over \delta
   (\partial^\kappa  A^\lambda_r)}
   = -  F_{\kappa\lambda}^r \ \  {\rm and} \quad
   {\delta  {\cal L} \over \delta  A^\lambda_r}
   = -  g  J_\lambda^r \ ,\ \ {\rm with}\quad
    J^\nu_a \equiv
    f^{ars} F^{\nu\kappa}_r A_\kappa^s
  + \overline \Psi \gamma^\nu  T^a \Psi \ .
\end {equation}
Canonical field theory yields straightforwardly
to the {\it color Dirac equations}
$ \bigl(i \gamma^\mu  {\cal D} _\mu -  m \bigr) \Psi  = 0$.
It also gives the {\it color Maxwell equations}, which are given
here in two conventions, {\it i.e.}\ %
\begin {eqnarray}
 \partial_\mu  {\bf F}^{\mu \nu} & = &  g {\bf J}^\nu \ ,
 \qquad\ {\rm with}\quad   {\bf J}^\nu \equiv
   -i\bigl[ {\bf F}^{\nu\kappa}, {\bf A}_\kappa\bigr]
  + \overline \Psi \gamma^\nu  T^a \Psi   T^a \ ,\\
    {\bf D}_\mu  {\bf F}^{\mu \nu} & = &  g \ {\bf J}_{\rm Q}^\nu \ ,
 \qquad\ {\rm with}\quad \ {\bf J}_{\rm Q}^\nu \equiv
  \overline \Psi \gamma^\nu  T^a \Psi   T^a \ \  {\rm and} \quad
    {\bf D}_\mu \equiv
   \partial_\mu + i g \bigl[ {\bf A}_\mu , \ast \bigr] \ .
\end {eqnarray}
The {\it color Maxwell current} ${\bf J}^\mu $ and the {\it quark
matter current} $\ {\bf J}_{\rm Q}^\mu $ have different conservation laws.
In particular
$ \partial_\mu {\bf J}^\mu = 0$.
The stress tensor
$ \Theta^{\mu\nu} =  F^{\mu\kappa}_a
   A_\kappa^a
  + {1\over2} \bigl[ \overline \Psi  i\gamma^\mu \partial^\nu \Psi
  + \ {\rm h.c.} \bigr]
 - g^{\mu\nu} {\cal L} $ is, at first,
is {\it not} manifestly gauge-invariant. But with the Maxwell
equations one derives
$ F^{\mu\kappa}_a \partial^\nu  A_\kappa^a =
  F^{\mu\kappa}_a  F^\nu_{\phantom{\nu} \kappa,a}
 + g  J^\mu_a  A^\nu_a
 + g  f^{ars} F^{\mu\kappa}_a  A^\nu_r  A_\kappa^s
  + \partial_\kappa\bigl(F^{\mu\kappa}_a  A^\nu_a \bigr)$,
and thus
\begin {equation}
 \Theta^{\mu\nu} = 2{\rm Tr} \bigl({\bf F}^{\mu\kappa}
   {\bf F}_\kappa^{\phantom{\kappa}\nu}\bigr)
  + {1\over2} \bigl[ \overline \Psi  i\gamma^\mu  {\cal D} ^\nu \Psi
  + \ {\rm h.c.} \bigr]
 - g^{\mu\nu} {\cal L}
  - 2\partial_\kappa
              {\rm Tr} \bigl({\bf F}^{\mu\kappa} {\bf A}^\nu \bigr) \ .
\end {equation}
All explicit gauge dependence resides in the last term.
For periodic boundary conditions it vanishes upon integration.
The generalized momenta `on the light cone' become then
{\it manifestly gauge invariant}, {\it i.e.}\ %
\begin {equation}
    P^\nu \equiv \int_\Omega \! d\omega \ \Theta^{+\nu}
   = \int_\Omega \! d\omega \ \bigl(2{\rm Tr} ({\bf F}^{+\kappa}
    {\bf F}_\kappa^{\phantom{\kappa}\nu})
  - g^{+\nu} {\cal L}  + {1\over2}
 \bigl[ \overline \Psi  i\gamma^+  {\cal D} ^\nu \Psi
  + \ {\rm h.c.} \bigr] \bigr) \ .
\label{eGenMom}
\end {equation}
Integration goes over all space-like coordinates ($d\omega$) and
$\Omega$ denotes the integration volume. This was first shown in
\cite{JaM80}.
Note that all this holds rigorously for SU(N) in (d+1) dimensions.
%---+----1----+----2----+----3----+----4----+----5----+----6----+----7

\noindent
{\underline {Light-Cone Coordinates.}}
We follow the convention of Kogut and Soper \cite{KoS70},
in particular with $ x^\pm \equiv (x^0 \pm  x^1)/\sqrt{2}$.

\noindent
\underline{Color Helicity Basis.}
We define the color helicity basis for SU(2) by the
Pauli matrices $\sigma^a$:
\begin{equation}
  \tau ^3 = {1\over2} \sigma^3 \ , \quad
  \tau ^\pm \equiv {1\over{2\sqrt{2}}} (\sigma^1 \pm i \sigma^2)
\ .\end{equation}
We can turn this into a vector space by introducing
elements $x^a$ such that tilde quantities are defined
with respect to the helicity basis, and untilded
the usual Cartesian basis:
\begin{equation}
x^a = \left(\begin{array}{c}
		x^1 \\
		x^2 \\
		\end{array}
	\right)
\qquad {\rm {and}}\qquad
\tilde{x}^a = \left(\begin{array}{c}
		\tilde{x}^1 \\
		\tilde{x}^2 \\
		\end{array}
		\right)
	= \left(\begin{array}{c}
		x^+ \\
		x^- \\
	 	\end{array}
	\right).
\end{equation}
The relation between the tilde and untilde basis can be
written
$\tilde{x}^a = \Lambda^a_b x^b
\; {\rm{and}} \;
x^a = \tilde{\Lambda}^a_b \tilde{x}^b $
where $\tilde{\Lambda} = \Lambda^\dagger$.
With these elements we can construct the metric
in terms of the tilde basis. Essentially we must
demand the invariance of the inner product of
any two vector space elements,
$ x^a y_a = \tilde{x}^a \tilde{y}_a \;.$
Using the fact that the metric in the $a=1,2$
basis is just the Kronecker delta $\delta_{ab}$
and the transformed metric is
$ \tilde{{\cal G}}_{ab} =
\tilde{\Lambda}_a^c \delta_{cd} \tilde{\Lambda}^d_b $.
Thus
\begin{equation}
 \Lambda =  {1\over{\sqrt{2}}}
	\left(\begin{array}{rr}
		1 & i \\
		1 & -i \\
		\end{array}
	\right)
\quad {\rm{and}} \quad
\tilde{{\cal G}}_{ab} =
		\left(	\begin{array}{cc}
			0 & 1 \\
			1 & 0 \\
			\end{array}
		\right) \;.
\end{equation}
The metric to raise and lower indices
in the helicity basis becomes
$ x_{\pm} = x^{\mp} $.
The color algebra looks formally like
the Lorentzian structure in light-cone coordinates.

\section {Gribov copies, Centre Conjugations and Fock Space}
\label{sec:GribFock}
\noindent
{\underline{Gribov Copies:}}
Because
of the torus geometry of our space and the non-Abelian
structure of the gauge group, there remain large gauge
transformations which are still symmetries of the theory
\cite{Gri78,Sin78} despite our complete fixing of the theory with
respect to small gauge transformations.
These are generated by local SU(2) elements
\begin{equation}
	V (x^-) = \exp ({- i  n_0 \pi {x^- \over {L}}   \tau _3)},
\;  n_0 \; {\rm{an \; even \;integer}}
\label{gribcopy}
\end{equation}
which satisfy periodic boundary conditions. Another symmetry of the
theory is $Z_2$ centre symmetry which here means allowing for
antiperiodic $V$ or alternately $ n_0$ odd. In both cases one
preserves the periodic boundary conditions on the gauge
potentials.
On the diagonal component of $ A^+$ it generates shifts that
are best expressed in terms of the dimensionless $ z  $,
namely $ z \rightarrow  z ' =  z + n_0$.
On the scalar adjoint field and its momenta
the effect of the transformation is
\begin {eqnarray}
	 \varphi_3  & \rightarrow & \varphi_3 \quad {\rm{and}} \quad
	 \varphi_\pm \rightarrow  \varphi_\pm
		\exp{(\mp i  n_0 {\pi\over L} x^-)},
\label{gribfield} \\
 \pi^3  & \rightarrow & \pi^3 \quad {\rm{and}} \quad
 \pi^\pm \rightarrow \pi^\pm
		\exp{(\pm i  n_0 {\pi\over L} x^-)}.
\label{gribmom}
\end {eqnarray}

\noindent
{\underline{Color Property of $ z $:}}
We now show that the gauge mode
$ z $ can be written in terms of an explicitly
color singlet object, thus demonstrating that it
itself is a color singlet and a viable physical
degree of freedom.
We construct the Wilson line by a contour $ {\rm C} $ along
the $ x $ direction from $- L $ to $ L $
\begin{equation}
 W  = {\rm Tr} {\rm {P}} \exp (i g \int_{{\rm C}} dx_\mu
 {\bf A}^\mu) = {\rm Tr} {\rm {P}}
\exp (i g \int_{- L}^{+ L} d x   {\bf A}^+).
\end{equation}
In the gauge employed in this paper, this is simply
$  W  = {\rm Tr} \exp (2\, i\, z \,\pi \, \tau ^3)
= 2 \cos (2 \pi  z)
\; ,$
and one can relate $ z $ to $ W $ modulo the integers,
$ z  = {1\over{2\pi}} {\rm{arcos}} ({{W}\over2})
\;.$
The integer shifts are nothing but the Gribov copies
discussed earlier. Observe that the dynamical quantity
$ W $ attains its minimum value at $ z ={\textstyle{1\over2}}$
matching with the minimum in the Fock vacuum potential.
Since $ W $ is explicitly constructed in terms of
a color trace, $ z $ is a color singlet.
%---+----1----+----2----+----3----+----4----+----5----+----6----+----7

\section{The Charge Operators}
\label{sec:charges}
In the text we introduced the operators which are the
Discrete Fourier transforms of the scalar current components
with the constrained zero mode {\it removed}.
The explicit expressions for these in terms of the various
Fock operators are:
\begin {eqnarray}
   Q_3 (k) = - \sum_{n = {1\over2}}^\infty
                 \sum_{m = {1\over2}}^\infty
        b_m  d_n \ \Bigl({u_m \over  v_n}
                            - {v_n \over  u_m} \Bigr)
                     \delta^k_{n + m}
           & + & \sum_{n = {1\over2}}^\infty
               \sum_{m = {1\over2}}^\infty
        b^\dagger_n  b_m \ \Bigl({u_n \over  u_m}
                            + {u_m \over  u_n} \Bigr)
                     \delta^m_{n + k}
\nonumber \\
           & - & \sum_{n = {1\over2}}^\infty
               \sum_{m = {1\over2}}^\infty
        d^\dagger_n  d_m \ \Bigl({v_n \over  v_m}
                            + {v_m \over  v_n} \Bigr)
                     \delta^m_{n + k}    \ ,
\label{charge3} \\
   Q_+ (k) = + \sum_{n =  1}^\infty
                 \sum_{m = {1\over2}}^\infty
        a_n  d_m \ \Bigl({w_n \over  v_m}
                            - {v_m \over  w_n} \Bigr)
                     \delta^k_{n + m}
           & + & \sum_{n =  1}^\infty
               \sum_{m = {1\over2}}^\infty
        a^\dagger_n  d_m \ \Bigl({w_n \over  v_m}
                            + {v_m \over  w_n} \Bigr)
                     \delta^m_{n + k}
\nonumber \\
           & - & \sum_{n =  1}^\infty
               \sum_{m = {1\over2}}^\infty
        a_n  b^\dagger_m \ \Bigl({w_n \over  u_m}
                            + {u_m \over  w_n} \Bigr)
                     \delta^n_{m + k} \ ,
\label{charge+} \\
   Q_- (k) = + \sum_{n =  1}^\infty
                 \sum_{m = {1\over2}}^\infty
        a_n  b_m \ \Bigl({w_n \over  u_m}
                            - {u_m \over  w_n} \Bigr)
                     \delta^k_{n + m}
           & + & \sum_{n =  1}^\infty
               \sum_{m = {1\over2}}^\infty
        a^\dagger_n  b_m \ \Bigl({w_n \over  u_m}
                            + {u_m \over  w_n} \Bigr)
                     \delta^m_{n + k}
\nonumber \\
           & - & \sum_{n =  1}^\infty
               \sum_{m = {1\over2}}^\infty
        a_n  d^\dagger_m \ \Bigl({w_n \over  v_m}
                            + {v_m \over  w_n} \Bigr)
                     \delta^n_{m + k}     \ .
\label{charge-}
\end   {eqnarray}
%---+----1----+----2----+----3----+----4----+----5----+----6----+----7

\section{Analysis of the Gauge Potential}
\label{sec:gaugepot}
%---+----1----+----2----+----3----+----4----+----5----+----6----+----7
The gauge potential $ V_0 (z)$ was defined in Eq.(\ref{gaugepoteq}).
We express it here conveniently
\begin  {eqnarray}
   V_0 (z)  & = & \sum_{k = 1}^\infty  S_3(k)  w_k^4
  + \sum_{k ={1\over2}}^\infty \Bigl(
     S_+(k)  v_k^4 + S_-(k)  u_k^4 \Bigr)
\ , \\
\ {\rm with}\qquad   S_3(k , z)  & \equiv &
 \langle 0 \vert \Bigl[ Q_3 (k) , Q_3^\dagger (k) \Bigr]
 \vert 0\rangle = \sum_{n ={1\over2}}^\infty
                  \sum_{m ={1\over2}}^\infty
  \Bigl({u_m \over  v_n}
         - {v_n \over  u_m} \Bigr)^2 \delta^k_{m + n}
\ , \label{commS3} \\
  S_+(k, z)  & \equiv &
 \langle 0 \vert \Bigl[ Q_+ (k) , Q_+^\dagger (k) \Bigr]
 \vert 0\rangle = \sum_{n = 1}^\infty
                  \sum_{m ={1\over2}}^\infty
  \Bigl({w_n \over  v_m}
         - {v_m \over  w_n} \Bigr)^2 \delta^k_{m + n}
\ , \label{commS+} \\
      S_-(k, z)  & \equiv &
   \langle 0 \vert \Bigl[ Q_- (k) , Q_-^\dagger (k) \Bigr]
 \vert 0\rangle = \sum_{n = 1}^\infty
                  \sum_{m ={1\over2}}^\infty
  \Bigl({w_n \over  u_m}
         - {u_m \over  w_n} \Bigr)^2 \delta^k_{m + n}
\ , \label{commS-} \end    {eqnarray}
in terms of the commutator functions $ S (k)$. Thus
\begin  {equation}
  V_0 (z)  = \sum_{n ={1\over2}}^\infty
           \sum_{m ={1\over2}}^\infty
  \Bigl({u_m \over  v_n}
         - {v_n \over  u_m} \Bigr)^2  w^4_{m + n}
         + \sum_{n = 1}^\infty
         \sum_{m ={1\over2}}^\infty
  \Bigl({w_n \over  v_m}
         - {v_m \over  w_n} \Bigr)^2  v^4_{m + n}
  + \Bigl({w_n \over  u_m}
         - {u_m \over  w_n} \Bigr)^2  u^4_{m + n}
\ .\end    {equation}
All these functions depend on $ z $ through $\zeta (z)$
in the coefficients
$ u_n(z)=1/\sqrt{n+\zeta}$, $ v_n(z)=1/\sqrt{n-\zeta}$, and
$ w_n = 1/\sqrt{n}$.
The gauge potential is thus
{\it manifestly invariant} under large gauge transformations.
By inspection, it has {\it singularities} at all values
$ z \in \ Z $, particularly at $ z  = 0$ and $ z  = 1$.
Noting that the coefficients have the symmetry
$  u_m  (-\zeta) =  v_m  (\zeta) $ and
$  v_m  (-\zeta) =  u_m  (\zeta) $,
one observes another symmetry, namely
$  V_0 (-\zeta) =  V_0 (\zeta) $.
This implies that the gauge potential is symmetric around
$ z = {1\over2}$ in the fundamental modular domain,
having there a {\it minimum}.
It is thus reasonable to renormalize the gauge potential by
$  G  (z) =  V_0  (z) -  V_0  ({1\over2}) $.
One can expand $ V_0 (z)$ around its minimum at $ z ={1\over2}$.
The first derivative vanishes there,
and the second can be cast into the form
\begin  {eqnarray}
 V_0^{\prime\prime} (z)
  &=& \sum_{n={1\over2}}^s \sum_{m={1\over2}}^\infty \Bigl[
    {2\over (n-\zeta)    (m+\zeta)^3}
  + {2\over (n+\zeta)    (m-\zeta)^3}
  - {2\over (n-\zeta)^2 (m+\zeta)^2} \Bigr]
\nonumber \\
  &+& \sum_{n=1}^s \sum_{m={1\over2}}^\infty \Bigl[
     {2\over n  (m-\zeta)^3}
   + {2\over n  (m+\zeta)^3}
   - {24\over (n+m-\zeta)^4}
   - {24\over (n+m+\zeta)^4}  \Bigr] \ .
\end    {eqnarray}
The first two terms in each line diverge as $s \rightarrow \infty$.
A measure for the curvature at $\zeta ({1\over2})=0$
is obtained by means of Riemann's zeta function $\zeta $
(not to be confused with $\zeta (z)$), namely
\begin  {equation}
 \omega^2 \equiv {1\over 8}  V_0^{\prime\prime}({1\over2})
       = \ln{s} \sum_{n =1}^\infty {1 \over(n -{1\over2})^3}
       = 7 \zeta(3) \ln{s}
     \simeq 8.4144 \ln{s}
     \equiv \omega_0^2 \ln{s} \ .
\end    {equation}
We have thus analytically obtained the divergent $s$-dependent
part of the gauge potential.

Finally, we analyze the commutator functions at $ \zeta = 0 $.
Direct evaluation of Eqs.(\ref{commS3}) to (\ref{commS-})
yields the positive values
\begin  {equation}
  S_3   (k)  =
 \sum_{n ={1\over2}}^\infty \sum_{m ={1\over2}}^\infty
 \Bigl(\sqrt {{n \over  m}}
      - \sqrt {{m \over  n}} \Bigr)^2 \delta^k_{m + n}
\ , \quad {\rm and} \qquad
  S_\pm (k)  =
 \sum_{n = 1}^\infty \sum_{m ={1\over2}}^\infty
 \Bigl(\sqrt {{n \over  m}}
      - \sqrt {{m \over  n}} \Bigr)^2 \delta^k_{m + n}
\ . \label {cc3func} \end    {equation}
For sufficiently large values all of them
approach $ S (k) \simeq 2k \ln{k}$.
%---+----1----+----2----+----3----+----4----+----5----+----6----+----7
 \end   {appendix}
 \newpage
FIGURE CAPTION:

Fig.1. The gauge potential $ G (z) \equiv  V_0 (z) -  V_0 ({1\over2})$
in the fundamental modular domain depicted for the two scenarios:
(a) (full curve) with the cutoff dependence removed.
(b) (dashed curve) with the cutoff dependence kept. In the
latter case the cutoff $s = 131$ leading to the potential
$ V_0  (z  - {1\over2})^2$. In this oscillator well we give the
lowest energy eigenvalue $E_0 = 25.66$.

%---+----1----+----2----+----3----+----4----+----5----+----6----+----7
 \end   {document}